\documentclass{ws-procs9x6}

\begin{document}

\title{LORENTZ VIOLATION IN THE LINEARIZED GRAVITY}

\author{A.F.\ FERRARI}

\address{Universidade Federal do ABC, Centro de Ci\^encias Naturais e Humanas\\
Rua Santa Ad\'elia, 166, 09210-170, Santo Andr\'e, SP, Brazil\\
E-mail: alysson.ferrari@ufabc.edu.br}

\author{A.YU.\ PETROV}

\address{Departamento de F\'{\i}sica, Universidade Federal da Para\'{\i}ba\\
Caixa Postal 5008, 58051-970, Jo\~ao Pessoa, Para\'{\i}ba, Brazil\\
E-mail: petrov@fisica.ufpb.br}

\begin{abstract}
We study some consequences of the introduction of a Lorentz-violating
modification term in the linearized gravity, which leads to modified dispersion
relations for gravitational waves in the vacuum. We also discuss possible mechanisms for the 
induction of such a term in the Lagrangian. 
\end{abstract}

\bodymatter

\section{Introduction}
In recent years, a great amount of work has been done exploring the possibility
of very small departures of Lorentz invariance, the Standard-Model Extension proposed by Kosteleck\'y
and Colladay\cite{1} furnishing a general background for such investigations. 
One of the interesting possibilities arising in such a context is the modification of
the dispersion relations governing the propagation of particles in the vacuum,
which could generate outstanding effects in astrophysical observations, for instance\cite{2}. 
For gravity theories, the incorporation of Lorentz violation is delicate,
in particular when the violation is not spontaneous\cite{3}; nevertheless, it is worthwhile to investigate 
how to introduce Lorentz violation in gravity in such a way that simple physical 
effects can be studied. This is one of the objectives of our work. 

We consider the linearized gravity theory augmented
by one Lorentz violating term in the Lagrangian and show that it leads to a modified dispersion relation
for the propagation of gravitational waves in the vacuum. Next, we discuss
one possible mechanism for the generation of such a term, based on a 
deformation of the Poisson algebra of the metric fluctuation and its canonical conjugated momenta\cite{4}. 
More details of this calculation, together with another method to generate the particular form of Lorentz violation
we consider here, are discussed elsewhere\cite{5}.

\section{The modified Fierz-Pauli action}

The starting point of our study is the inclusion of a Lorentz-violating term $\Delta L$,
\begin{equation}
\label{newterm}
\Delta L\,=\,-2\epsilon^{\lambda\mu\nu\rho}\theta_{\rho}
h_{\nu\sigma}\partial_{\lambda}h_{\mu}^{\sigma}\,,
\end{equation}
in the Einstein-Hilbert Lagrangian in the weak field approximation, also known as the Fierz-Pauli Lagrangian.
Here, $h_{\mu\nu}$ is the metric fluctuation, and $\theta^{\rho}=(0,\theta^i)$, $i=1,2,3$ is 
a parameter for the Lorentz violation. 
The term $\Delta L$ breaks the gauge invariance of the model 
(which is the linearized version of the original diffeomorphism invariance of General Relativity).
Since there exists a preferred direction in 
spacetime, the angular momentum is no longer
conserved, however, the energy-momentum tensor is still conserved 
since the background (Minkowski) spacetime is homogeneous. 
As for the Bianchi identities, they are not satisfied, 
and this problem can be traced back to the
general incompatibility between explicit Lorentz violation
and Riemann-Cartan geometry\cite{3}.

To study gravitational wave propagation, we show that the equations of motion for the 
traceless transversal part
of the metric fluctuation ${\tilde h}_{\mu\nu}$ are gauge invariant and decouple from the other componentes 
of ${h}_{\mu\nu}$; this happens if $\theta_{ik}\partial_k
{\tilde h}_{ij}  =  0$. If this condition holds, we end up with 
\begin{equation}
\label{eomh}
\frac{1}{2}\Box \tilde h_{ij} + 2 \left[ 
\theta_{ik}\dot{\tilde h}_{kj} + \theta_{jk}\dot{\tilde h}_{ki}  \right]
\, = \, 0 \,.
\end{equation}
Here, we defined an antisymmetric symbol $\theta_{ij}=-\epsilon_{0ijk}\theta^k$.

The above-mentioned condition can be met if we choose $\theta^i = ( 0, 0, \theta / 4 )$, such that the only
nonvanishing $\theta_{ij}$ are $\theta_{12}=-\theta_{21}=-\theta/4$, and consider a
wave propagating in the $x_3$ direction. In this case, Eq.\ \eqref{eomh} can be
solved by the ansatz $\tilde{h}_{ij} = H_{ij}e^{iq^\mu x_\mu}$, $q = (E, \vec
p)$, and we end up with only two independent equations,
$\Box Z \pm 2i\theta\dot{Z}=0$,
where $Z=H_{11}-iH_{12}$, $\bar{Z}=H_{11}+iH_{12}$. The corresponding dispersion
relations are given respectively by
$E=-\theta\pm\sqrt{{p}^2+\theta^2}$ and $E=+\theta\pm\sqrt{{p}^2+\theta^2}$,
with $p=|\vec{p}|$. Thus the dispersion relations are
modified, exhibiting a kind of birefringence phenomenon for the two circular polarizations of the 
gravitational waves.

\section{The generation of $\Delta L$}

One possibility for the induction of Eq.\ \eqref{newterm} is
based on the deformation of the Poisson algebra of the
canonical variables\cite{4,7}. The inspection of the constraint structure of the Fierz-Pauli
action yields a set of primary constraints $\Phi_\mu
^{(1)}=p_{0\mu}\simeq 0$, and secondary ones $
\Phi^{(2)}_j \, = \, \partial_i p_{ij}\simeq 0 $ and $\Phi^{(2)}_0 \, = \, \partial_l\partial_l h_{kk}-\partial_i\partial_j h_{ij}\simeq 0 $, which act
as generators of the gauge symmetry.

The proposed deformation consists of assuming the noncommutativity of the canonical conjugated momenta,
$\left\{p_{ij}(\vec{x}),p_{kl}(\vec{y})\right\}=\theta_{ijkl}\delta(\vec{x}-\vec{
y})$,
where $\theta_{ijkl}$ is a symbol possessing symmetry
$\theta_{1234}=\theta_{2134}=\theta_{1243}=-\theta_{3412}$. 
We then require that, within this modified algebra,
the secondary constraints still generates the same gauge symmetries of the
undeformed theory. To this end, we
have to modify the secondary constraints as follows,
\begin{equation}
\label{modgen}
R_k \, = \, \partial_i p_{ik}-\theta_{klnm}\partial_l h_{nm} \, .
\end{equation}
This modification implies a modification of the Hamiltonian of the model, and
therefore of the Lagrangian, which turns out to be
\begin{equation}
\label{lagrnew}
L_{\rm  new} = L_{\rm FP} - \left[
2\theta_{jlnm}\partial_l h_{nm} +\frac{1}{2}\dot{h}_{ij}\theta_{klij}h_{kl} \right]
\,.
\end{equation}
To verify whether this procedure can reproduce $\Delta L$, we make an assumption
on the form of the $ \theta_{jlnm} $. The most appropriate choice seems to be
\begin{equation}
\theta_{ijkl}=\theta_{ik}\tilde{\delta}_{jl}+\theta_{il}\tilde{\delta}_{jk}
+\theta_{jl}\tilde{\delta}_{ik}+\theta_{jk}\tilde{\delta}_{il},
\end{equation}
where $\tilde{\delta}_{ij}=\delta_{ij}-{\partial_i\partial_j}/{\nabla^2}$. The induced
term in the classical action is
\begin{equation}
\Delta
L=2\theta_{ki}\dot{h}_{ij}\left(\delta_{lj}-\frac{\partial_l\partial_j}{\nabla^2}\right)h_
{kl}
-4h_{0j}\theta_{ln}\left(\delta_{jm}-\frac{\partial_j\partial_m}{\nabla^2}\right)\partial_lh_{
nm}.
\end{equation}
This term is invariant under gauge transformations satisfying
$(\delta_{jm}-{\partial_j\partial_m}/{\nabla^2})\xi_m=0$. This (restricted) gauge
freedom can be used to fix $\partial_m h_{mn}=0$ and $h_{0m}=0$, thus obtaining the
simple modification $\Delta L=2\theta_{ki}\dot{h}_{ij}h_{kj}$,
which leads to the same modified dispersion relations we obtained before. Even
if this deformation does not reproduce exactly the term $\Delta L$, it leads to 
a modified gravity theory with (restricted) gauge invariance, 
and with the same deformed dispersion relations. 

We notice that this is not the only Lorentz-violating additive term which can be obtained with use of 
the deformation of the Poisson algebra. As an example, if we assume 
$\tilde{\delta}_{ij}=\delta_{ij}\nabla^2-\partial_i\partial_j$, 
we find that the following gauge-invariant term is generated in the action,
\begin{eqnarray}
\label{delta3}
\Delta L&=&-\theta^p 
\left[ 2\epsilon_{0kip}\dot{h}_{ij}\left(\delta_{lj}\nabla^2-\partial_l\partial_j\right)h_{kl}\right.
\nonumber\\&&
\left.-4h_{0j}\epsilon_{0lnp}\left(\delta_{jm}\nabla^2-\partial_j\partial_m\right)\partial_lh_{nm}
\right],
\end{eqnarray}
where $\theta_{ij}=-\epsilon_{0ijk}\theta^k$.

One may compare this with the well-known gravitational Chern-Simons term \cite{Jackiw},
$\Delta L^{CS}=2\theta^{\lambda}h^{\mu\nu}\epsilon_{\alpha\mu\lambda\rho}\partial^{\rho}
[\Box h^{\alpha}_{\nu}-\partial_{\nu}\partial_{\gamma}h^{\gamma\alpha}]$;
if the $\theta^\lambda$ vector is chosen to be purely spacelike, and the axial gauge $h_{0m}=0$ is imposed, 
the gravitational Chern-Simons term is reduced to
\begin{eqnarray}
\Delta L^{CS}=-2\theta^l\dot{h}^{ij}\epsilon_{lki0}[\Box h^k_j-\partial_j\partial_sh^{ks}],
\end{eqnarray}
which, under the gauge conditions $\partial_sh^{ks}=0$ and $h_{0i}=0$, is almost 
equivalent to Eq.\ \eqref{delta3}, up to the term with third time derivative.

\section{Summary}

We have studied the consequences of a Lorentz-violating term in the propagation of gravitational waves in the linearized gravitation
theory, finding a modified dispersion relation with a kind of birefringence effect. 
We have also shown how to generate this term via an appropriate deformation of the canonical algebra of the model, generalizing some
ideas that have been proposed in connection with noncommutative gauge theories before.

\section*{Acknowledgments}
This work is partially supported by the
Brazilian agencies CNPq, FAPESP and CAPES.
The work by A.Yu.\ P.\ is supported by the CNPq project 303461-2009/8.

\end{document}